\documentclass{article}

\usepackage{arxiv}

\usepackage[utf8]{inputenc} 
\usepackage[T1]{fontenc}    
\usepackage{hyperref}       
\usepackage{url}            
\usepackage{booktabs}       
\usepackage{amsfonts}       
\usepackage{nicefrac}       
\usepackage{microtype}      
\usepackage{lipsum}

\usepackage{chngcntr}\usepackage{graphicx}
  
\usepackage{dcolumn}
\usepackage{stackengine}
\usepackage[section]{placeins}
\usepackage[dvipsnames*,svgnames]{xcolor}
\usepackage{verbatim}
\usepackage[position=bottom,caption=false,captionskip=3pt,font=normalsize,subrefformat=simple,labelformat=simple]{subfig}

\usepackage{listings}
\usepackage{mathtools}
\usepackage{todonotes}
\usepackage{amsmath}
\usepackage{chngcntr}
\usepackage{bm}
\usepackage{algorithm}
\usepackage{algpseudocode}

\usepackage[super]{nth}
\usepackage{fixltx2e}
\usepackage[euler]{textgreek}

\usepackage{caption}

\hypersetup{
    colorlinks=true,
    linkcolor={blue},
    filecolor={blue},      
    urlcolor={blue},
    citecolor={blue}
}

\title{Mitigating the Effect of Nanoscale Porosity on Thermoelectric Power Factor of Si}

\author{
  S. Aria Hosseini, P. Alex Greaney\thanks{corresponding author: agreaney@engr.ucr.edu}\\
  Department of Mechanical Engineering\\ 
  University of California, Riverside\\
  Riverside, CA 92521, USA\\
   \And
 Giuseppe Romano\\
  Department of Mechanical Engineering\\
  Massachusetts Institute of Technology\\
  Cambridge, MA 02139, USA \\
}

\begin{document}
\maketitle

\begin{abstract}
The addition of porosity to thermoelectric materials can significantly increase the figure of merit, ZT, by reducing the thermal conductivity. Unfortunately, porosity is also detrimental to the thermoelectric power factor in the numerator of the figure of merit ZT. In this manuscript we derive strategies to recoup electrical performance in nanoporous Si by fine tuning the carrier concentration and through judicious design of the pore size and shape so as to provide energy selective electron filtering. In this study, we considered phosphorus doped silicon containing discrete pores that are either spheres, cylinders, cubes, or triangular prisms. The effects from these pores are compared with those from extended pores with circular, square and triangular cross sectional shape, and infinite length perpendicular to the electrical current. A semiclassical Boltzmann transport equation is used to model Si thermoelectric power factor. This model reveals three key results: The largest enhancement in Seebeck coefficient occurs with cubic pores. The fractional improvement is about 15\% at low carrier concentration ($< 10^{20}\ \mathrm{1/cm^3}$) up to 60\% at high carrier population with characteristic length around $\sim 1\ \mathrm{nm}$. To obtain the best energy filtering effect at room temperature, nanoporous Si needs to be doped to higher carrier concentration than is optimal for bulk Si. Finally, in \textit{n}-type Si thermoelectrics the electron filtering effect that can be generated with nanoscale porosity is significantly lower than the ideal filtering effect; nevertheless, the enhancement in the Seebeck coefficient that can be obtained is large enough to offset the reduction in electrical conductivity caused by porosity. \href{https://dx.doi.org/10.1021/acsaem.0c02640}{DOI:https://dx.doi.org/10.1021/acsaem.0c02640}
\end{abstract}

\keywords{bulk thermoelectric, electron transport, nanoporous, electron filtering}

\section{INTRODUCTION}

In the quest to create inexpensive thermoelectric (TE) materials that can be used for harvesting low grade waste heat, researchers have identified the strategy for improving thermoelectric performance in materials such as Si by engineering nanoscale porosity \cite{tang2010holey,lee2008nanoporous}. Nanoscale pores with a spacing smaller than the typical phonon mean free path hinder heat transfer by phonons and can produce a dramatic reduction in thermal conductivity \cite{romano2014toward}. However, although the electron mean free path is much smaller than the phonon mean free path, porosity damages the transport of low energy electrons. In this manuscript we examine strategies for designing pores so that the damage that they cause to the electrical conductivity is offset in the thermoelectric power factor (PF) by an enhancement in Seebeck coefficient. These materials are envisioned for conversion of waste heat, and so work in the temperature range of 300–700 K. Theoretical predictions suggest that at these temperatures there is significant scope for improving a materials’ thermoelectric power factor through electron energy filtering, thus our approach to mitigate the effect of pores is to identify conditions under which the reduction in electrical conductivity from additional scattering of electrons by pores is offset by improvement of the Seebeck coefficient due to electron energy filtering.

The performance of thermoelectric materials at a given temperature, $T$, is quantified by the dimensionless figure of merit, $ZT=(\sigma S^2)/(\kappa_e+\kappa_l)T$, where $\sigma$ is electrical conductivity, $S$ is Seebeck coefficient, $\kappa_e$ and $\kappa_l$ are electron and lattice thermal conductivity, respectively \cite{lee2010effects}. The TE figure of merit depends on a combination of strongly interdependent electrical transport properties, that have countervailing dependence on the carrier concentration so that the overall scope for enhancing the power factor is limited. The tradeoff of these parameters is well studied, and it has become an accepted truth that optimal performance of bulk TE can be obtained in semiconductors that are highly doped to a narrow window of optimized charge carrier concentration \cite{lee2010effects}.

The electrical transport properties that appear in ZT can be derived from the semiclassical Boltzmann transport equation using the single relaxation time approximation \cite{chen2005nanoscale}. In this model, the electrical conductivity, $\sigma$, is written as

\begin{equation}\label{eq:sigma}
    \sigma = -\dfrac{1}{3}e^2\int\chi(E,T)\tau(E,T)dE,
\end{equation}
where e is electron charge, $\tau(E,T)$ is momentum relaxation time of electrons with energy $E$ at temperature $T$ in \textit{n}-doped semiconductors. The kernel $\chi$ includes all the intrinsic non-scattering terms and is given by 

\begin{equation}\label{eq:chi}
\chi(E,T)=\nu^2(E)D(E)\dfrac{df(E,E_f,T)}{dE}.
\end{equation}
Here $E_f$ is the Fermi level, $\nu(E)$ the carrier group velocity, $f(E_f,E,T)$ the Fermi-Dirac distribution, and $D(E)$ is the density of states available for charge carriers. The Seebeck coefficient, $S$, in ZT describes the diffusion of electrons due to temperature gradient and is related to the difference between the average energy at which current flows and the Fermi energy level \cite{lundstrom2012near}. In bulk material, with negative charge carrier, the Seebeck coefficient is given by \cite{lundstrom2012near}

\begin{equation}\label{eq:chi}
S= \left( -\frac{k_B}{e}\right) \left(\frac{E_c-E_f}{k_BT}+\delta \right),
\end{equation}
where $k_B$, $E_c$ are Boltzmann constant and conduction band edge, respectively. The dimensionless parameter $\delta$ describes how far the average energy of the current carrying electrons is from the conduction band edge. It is defined as $\delta=\Delta_1/k_B T$, where $\Delta_1 = E_{\sigma}-E_c$, and $E_{\sigma}$ is the average energy of the charge carrier weighted by their contribution to electrical conductivity

\begin{equation}\label{eq:chi}
\Delta_n = \frac{\int \chi(E,T)\tau(E,T)E^ndE}{\int \chi(E,T)\tau(E,T)dE}.
\end{equation}

\begin{figure}[h]
\centering
\includegraphics[width=0.5\textwidth]{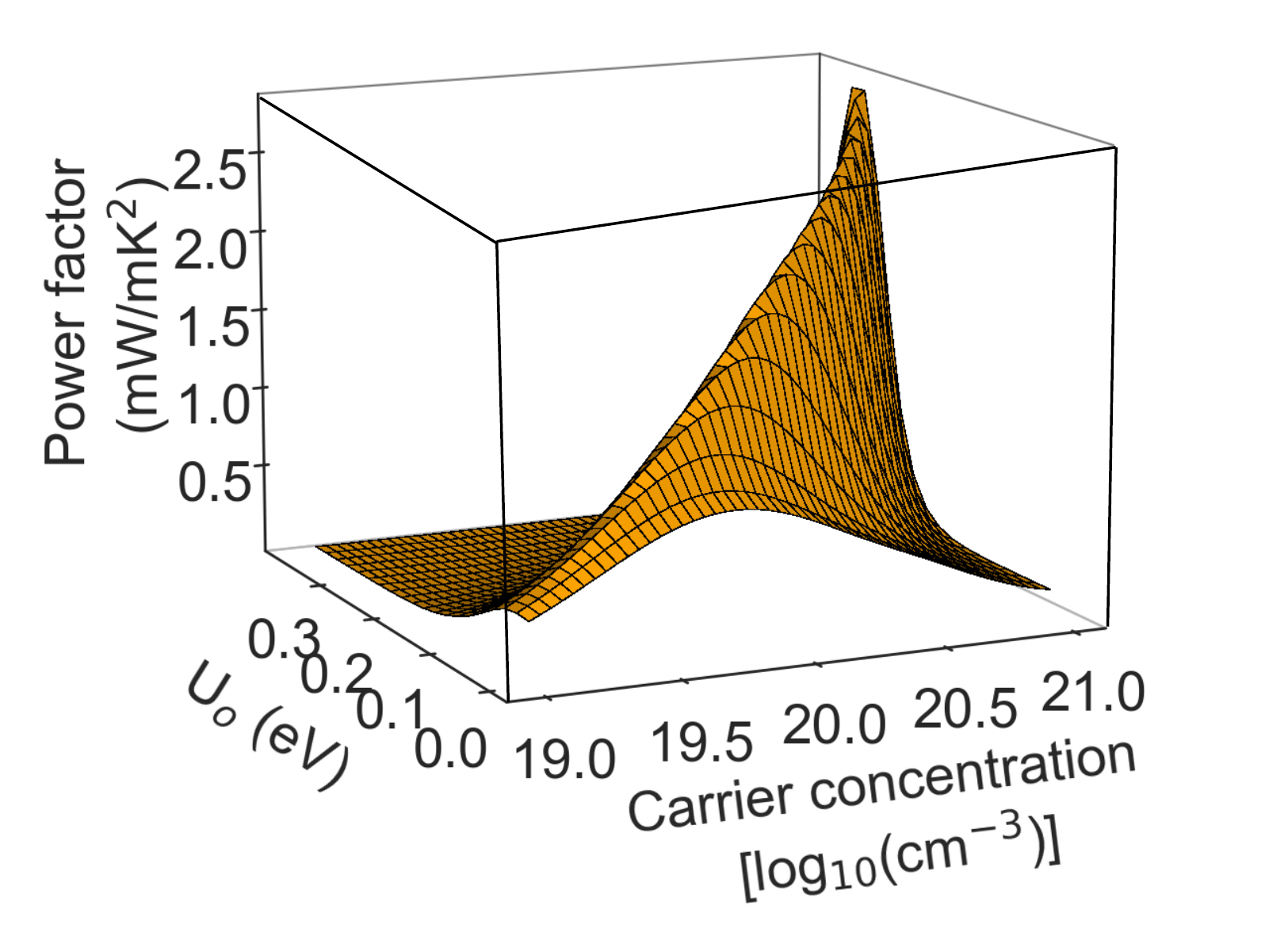}
\caption{Ideal energy filtering effect in bulk P-doped silicon at 300 K. This model predicts best power performance at the tail of the Fermi distribution.}
\label{fig:fig1}
\end{figure}

The central concept of energy filtering is to provide sources of scattering that selectively impede low energy electrons so as to increase $\Delta_1$  by reshaping product $\chi(E,T)\tau(E,T)$ so that it is more strongly asymmetric about the fermi energy. For ideal or perfect filtering a high rate of additional scattering would be applied to all the electrons with energy lower than a certain threshold, $U_o$ so as to reduce their drift velocity to zero. The calculated change in the room temperature power factor ($\sigma S^2$) of \textit{n}-doped silicon that would be provided by with ideal filtering is plotted in figure \ref{fig:fig1} as a function of filtering threshold, $U_o$, and carrier concentration (the details of this calculation are explained in the next section). The key result of this calculation is that if one can control the filtering threshold, the best power performance is to be found at high carrier concentration — ideal filtering breaks the conventional wisdom that there is a carrier concentration that provides the best compromise between conductivity and thermopower to optimize the power factor. It would provide game changing scope for enhancing thermoelectric power factor by exploiting the carrier population in the tail of the Fermi distribution. In this manuscript we examine the electron energy filtering effect provided in Si by nanoscale pores of various sizes and shapes. Our study shows while filtering by nanoscale pores are far from the ideal model, they can provide sufficient enhancement in Seebeck to countervail the degraded electrical conductivity, leaving power factor undiminished. In the sections that follow we describe electron scattering from discrete pores with different shapes, and the parameter-free semiclassical model that we use to model \textit{n}-type Si (and its validation). Then we elucidate the effect of pores that are extended in one dimension through the entire system perpendicular to transport direction. We conclude the manuscript with a brief discussion of Lorenz number and TE performance at high temperature.

\subsection{Electron Transport in Nanoporous Silicon}

We used semiclassical BTE models of electrical conductivity and Seebeck coefficient to predict the strength of the filtering effect in \textit{n}-type Si based nanoporous materials. This model is based on the intrinsic electronic band structure of undoped Silicon obtained using density functional theory (DFT). Without porosity, the scattering lifetime, $\tau$, is dominated by electron-ion and electron-phonon scattering, and the scattering rate from these two processes were modeled using equations in refs \cite{lundstrom2009fundamentals} and \cite{ravich1971scattering}, respectively. The Fermi level for a given carrier concentration was computed relative to the conduction band edge and self-consistently with the DFT band structure. The electron-pore scattering term was computed using Fermi’s golden rule and Matthiessen’s rule was used to add the scattering rates. The detail of the calculations for bulk Si can be found in \cite{Hosseini2021}, here we briefly explain the arc of the model.

The terms $D(E)$, and $\nu(E)$, in function $\chi$ for Si were derived from the conduction band of Si computed with DFT using the Vienna Ab initio Simulation Package (VASP) \cite{kresse1993ab,kresse1994ab,kresse1996efficiency,kresse1996efficient} using generalized gradient approximation (GGA) with the Perdew-Burke-Erzerhof exchange correlation functional (PBE) \cite{perdew1996generalized}. Projector augmented wave (PAW) pseudopotentials is used represent the ion cores \cite{blochl1994projector,kresse1999ultrasoft}. The Kohm-Sham wave functions constructed using a planewave basis set with 700 eV energy cutoff. The Brillouin zone was sampled using $12\times12\times12$ Monkhorst-Pack k-point grid \cite{monkhorst1976special}. The forces on the atoms minimized to better than $10^{-6}$ eV/Å to relax the Si primitive cell. The electronic band structure used to compute D(E) on a $45\times45\times45$ k-point grid. The group velocity was obtained from the conduction band curvature, $\nu = \frac{1}{\hbar} \left | \nabla_k E  \right |$ along the $\left\langle 100\right\rangle$ directions on the $\Gamma$ to $X$ Brillouin zone path. 

To complete the transport model, we need to define the electron lifetime, $\tau(E,T)$ in bulk Si — that is the coherence time of electrons between scattering events in bulk Si containing no pores. At moderate temperatures (room temperature), this scattering is predominated by a combination of electron-phonon and electron-ion interactions. Semiconductor TEs are generally doped to beyond the solid solubility limit (they are supersaturated) so that the carrier population is high, and the Coulomb potential is strongly screened. Therefore, we used the model developed for ions with strong screening for electron-ion lifetime \cite{lundstrom2009fundamentals}. In this model the transition rate has $\delta$-function form and the screening length plays a significant role. For better prediction of screening length, we used the generalized model for degenerate semiconductors \cite{mondal1986effect}. We used phonon deformation potential of Ravich to model electron-phonon lifetime \cite{ravich1971scattering}.

The final material property that appears in function $\chi$ is the Fermi level. In P-doped silicon this depends strongly on the carrier concentration, which varies non-monotonically with temperature as the solubility of the dopant changes. For a given carrier concentration, we used a self-consistent approach to compute $E_f$ by setting the conduction band edge as the reference frame and computing $E_f$ that gives the same carrier population in DFT computed band to circumvent the well documented problem of DFT’s underprediction of electronic band gaps.

\begin{figure}[h]
\centering
\includegraphics[width=0.5\textwidth]{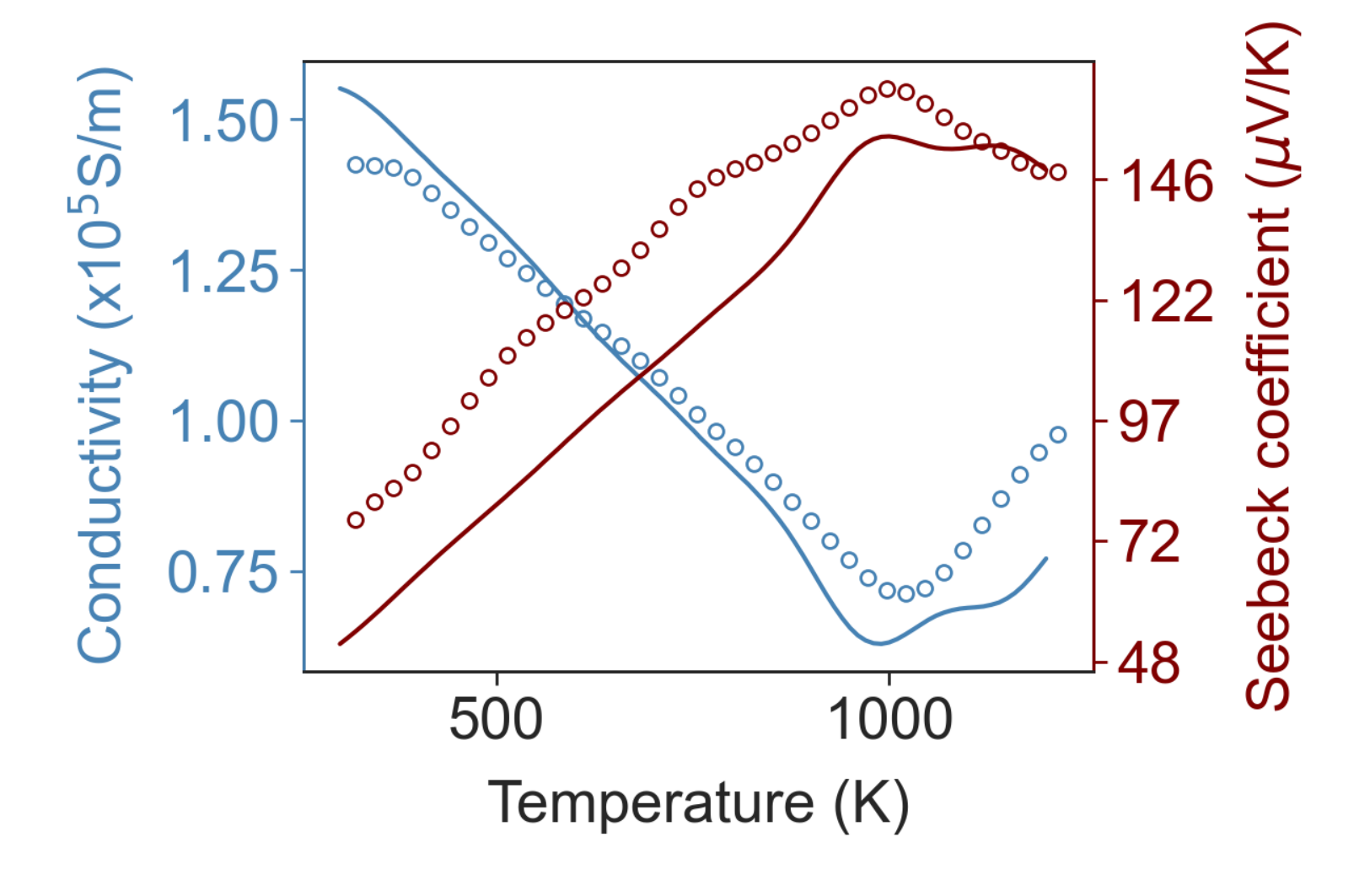}
\caption{Magnitude of electrical conductivity and Seebeck coefficient in phosphorus-doped bulk silicon. The solid blue line shows the model prediction for electrical conductivity, and the red line shows the prediction for the Seebeck coefficient. The experimentally measured $\sigma$ and $S$ are marked with open circles.}
\label{fig:fig2}
\end{figure}

We have validated the transport model in bulk materials against a set of phosphorus-doped Si based thermoelectrics produced through a novel plasma synthesis process whose synthesis and characterization are described in ref. \cite{Hosseini2021} . This reference also provides a complete description of the calculations of the electron lifetimes. The electrical conductivity and Seebeck coefficient of bulk P-doped Si is shown in figure \ref{fig:fig2}. The experimentally measured values are marked with open circles. The minimum (maximum) in conductivity (Seebeck coefficient) is due changes in the dissolved P with temperature. Our semiclassical model has no tuning parameters and uses the experimentally measured carrier concentration at each temperature as its only input. Its predictions for conductivity and thermopower are plotted in figure \ref{fig:fig2} using solid lines and are a good fit to the experimental data across the full range of temperatures. The calculations were performed used a python package, \textit{thermoelectric.py}, that we have made available for download through GitHub \cite{hosseini2019}.

The band bending in the Si at the interface to a pore presents a large potential energy barrier to electron transport. The height of this barrier, $U_o$, is equal to the semiconductor’s electron affinity \cite{lee2010effects}. The potential impedes transport of the low energy electrons while presenting little extra resistance to electrons in high energy states. This scattering, which occurs in addition to the intrinsic scattering from phonons and impurities, changes the electron lifetime by introducing a perturbation potential that for a single pore can be described as $U=U_o \Pi(r)$, where $\Pi(r)$ is a dimensionless boxcar function equal to unity inside the pore and zero outside of it. For uniform distribution of pores, the electron momentum relaxation time is defined as \cite{lee2010effects}

\begin{equation}\label{eq:taunp}
\tau_{np}^{-1}(s) = \frac{N}{8\pi^3} \int SR_{kk'} (1-\cos(\theta_{kk'})) dk',
\end{equation}
where $N$, is the number density of pores. This is related to porosity through $N=\varphi/V_{pore}$, where $\varphi$ is the porosity and $V_{pore}$ is the average pore volume. The term $SR_{kk'}$ in equation \ref{eq:taunp} is the probability of transition from an initial state with wave vector $k$ and energy $E$ to a state $k'$ with energy $E'$. The $(1-\cos \theta)$ term accounts for the change in momentum that accompanies this transition, with $\theta$ the angle between initial and scattered wavevectors. For a time-invariant potential, the transition rate $SR_{kk'}$ is given by Fermi’s golden rule \cite{lee2010effects,minnich2009modeling}, $SR_{kk'}=\frac{2\pi}{\hbar}\left|M_{kk'}\right|^2\delta(E-E')$. In this expression $M_{kk'}$ is the matrix element operator that describes the strength of the coupling between initial and final states and the number of ways that the transition between states can occur. For the Bloch waves, $M_{kk'}$ is defined as

\begin{equation}\label{eq:M}
M_{kk'}= \int e^{i(k'-k).r} U(r)dr.
\end{equation}

For energy conservative (elastic) electron-pore scattering only transmission to eigenstates with the same energy level is possible so the Brillouin zone integral in equation \ref{eq:taunp} can be written as a surface integral over the isoenergetic $k$ space contour

\begin{equation}
\label{eq:taunp2}
\tau_{np}^{-1}(s) = \frac{N}{(2\pi)^2\hbar}\int_{E(k')=E(k)}\frac{M_{kk'}\overline{M}_{kk'}}{\nabla E(k')}(1-\cos\theta)dS(k'),
\end{equation}
where $S(k')$ is the electron isoenergy state for a given wavevector. In most semiconductors isoenergy states close to the conduction valley have ellipsoid shape in momentum space that can be approximated as $E(k)=\hbar^2/2 \left[(k_x-k_{ox})^2/(m_x^* )+(k_y-k_{oy})^2/(m_y^* )+(k_z-k_{oz})^2/(m_z^*)\right]$, where $E(k)$, $k_o=(k_{ox},k_{oy},k_{oz})$, $m_x^*$, $m_y^*$, $m_z^*$ are energy level from conduction band edge, conduction band minimum, effective masses along $k_x$, $k_y$ and $k_z$, respectively. For Silicon the conduction band minimum is located at $k_o = 2\pi/a(0.85,0,0)$, where $a$ is the lattice parameter equal to $5.43\ \mbox{\normalfont\AA}$, and $m_x^*= 0.98\ m_o$, $m_y^*=m_z^*=0.19\ m_o$ where $m_o$ is electron rest mass equal to $9.11 \times 10^{-31}\ \mathrm{kg}$ \cite{sze2006pn}. We remark that in Si with narrow pore spacing confinement effect leads to flattening of conduction band \cite{cruz1996morphological, shi2015electronic} and increase the effect mass \cite{cruz1996morphological}, making transport coefficients different from the bulk Si. To avoid this regime, we limited our model to only consider low porosity within which the pores are far apart so that can be considered as perturbations encountered by the electronic wavefunctions of bulk Si. 

In this study, we considered phosphorus doped silicon containing one of four different shaped pores: spheres, cylinders, cubes, and triangular prisms, which are shown in the inset in the right-hand plot of figure \ref{fig:fig3}. The spherical pore has radius $r_o$, the edges of the cube are length $l_o$, the cylinder has radius $r_o$ and height $2r_o$ and all the edges of the triangular prism are $l_o$. The characteristic length (volume to surface area ratio) of these shapes are $1/3 r_o$, $1/6 l_o$, $1/3 r_o$ and $1/(2+4\sqrt{3}) l_o$, respectively. The analytic expression for the scattering matrix element, $M_{kk'}$ for each pore shape is presented in the appendix. The data for different pores’ geometries is plotted using the following markers: circles for spherical pores, squares for cubic pores, triangles for triangular prism pores and $\times$ markers for cylindrical pores. In spite of the poor thermoelectric efficiency of bulk silicon due to its high thermal conductivity, it provides an excellent platform for studying the role of design parameters on transport properties, since its bulk properties are extremely well characterized \cite{sze2006pn,shanks1963thermal,harter2019prediction}.

Introducing pores into Si will not change the concentration of carrier concentration locally in the remaining Si (nor the Fermi energy), but it will change the volume averaged carrier concentration due to the reduction in the volume averaged density of states. This will impact the conductivity, and thus the effective electrical conductivity of porous materials is modeled as $\sigma_{eff}=(1-\varphi) \sigma_{np}$. This change does not affect the Seebeck coefficient since the changes in density of states cancels out for the denominator and numerator of $S$ equation. We assumed that pores do not change the band structure of the Si — no quantum confinement effect, e.g., band flattening is considered — so we limit our study to the pores taking up 5\% volume fraction — a level that is still sufficient to reduce the thermal conductivity of Si by an order of magnitude \cite{romano2014toward}.

We assume that electron-pore scattering is independent of the electron-phonon and electron-ion scatterings thus Matthiessen rule can be used to sum the scattering rate from the three processes giving total scattering rate, $\tau^{-1}= \tau_b^{-1}+ \tau_{np}^{-1}$, where $\tau_b$ is the electron lifetime in bulk Si because of the ionic and phononic scattering terms ($\tau_b^{-1}= \tau_{ion}^{-1}+ \tau_{phonon}^{-1}$). 

\section{RESULTS AND DISCUSSION}

In this section we show the model prediction for the mean time between electrons being scattered by pores and demonstrate how this scattering changes the electrical behavior of nanoporous Si. The maximum power factor enhancement that can be obtained from electron filtering from nanopores is compared with that from ideal filtering. The electron scattering lifetimes due extended pores are also computed and it is shown that the electrical coefficients are insensitive to this class of pores due to limited unoccupied energy states. We conclude this section with a brief discussion of the effect of pores on electron thermal conductivity and the TE behavior of porous Si at high temperatures.

\begin{figure}[h]
\centering
\includegraphics[width=1\textwidth]{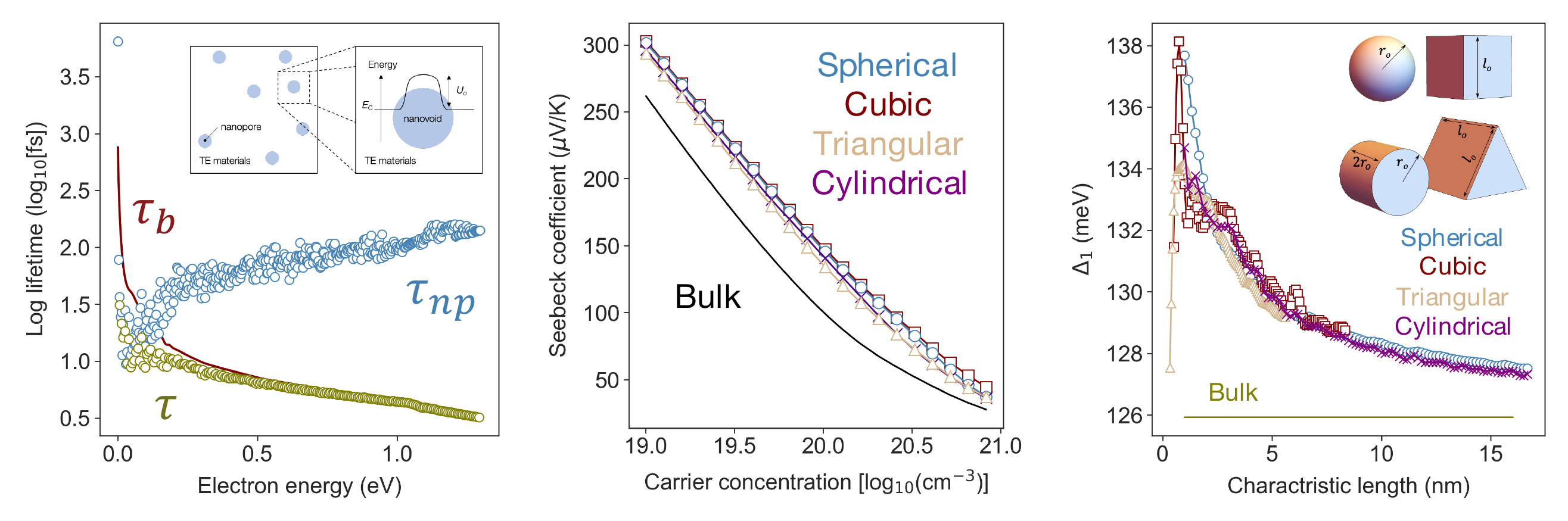}
\caption{(Left) Electron lifetime in \textit{n}-type Si at 300 K with a $10^{20}\ \mathrm{1/cm^3}$ carrier concentration due to: (red) scattering from phonon and ions, (blue) scattering from 0.05 porosity due to spherical pores with an 8 nm diameter, and (green) the resulting total lifetime from the combination of these processes. Electron-pore scattering is dominant for electron with energy less than 140 meV. The average energy of charge flow is increased about 10 meV by adding these pores. (Middle) The maximum Seebeck coefficients with different shapes of pores. The cubic and spherical pores show the best performance, respectively. (Right) The variation of average energy of electron vs effective length for pores with different shapes but fixed porosity of $\varphi = 0.05$ at $1.2 \times 10^{20}\ \mathrm{1/cm^3}$ concentration. The inset figure shows the four geometries examined in this work and their characteristic lengths in parenthesis. Clockwise from the top left: sphere ($l_c = 1/3r_o$), cube ($l_c = 1/6l_o$), triangular prism ($l_c = 1/(2 + 4\sqrt{3})l_o$), and cylinder ($l_c = 1/3r_o$).}
\label{fig:fig3}
\end{figure}

The left-hand plot in figure \ref{fig:fig3} shows the lifetimes in bulk Si because of the intrinsic electron-phonon and extrinsic electron-ion scattering (at $10^{20}\ \mathrm{1/cm^3}$ carrier population) in red and extrinsic electron-pore scattering computed in P-doped Si containing 5\% porosity due to 8 nm diameter spherical pores at 300 K in blue. The total lifetime is plotted in green. The noise in the lifetime reflects the difference in scattering rate of wavevectors around the conduction band valley minimum. Pores are the dominant scattering term for electrons with energy less than 140 meV. The central panel shows the Seebeck coefficient at different carrier concentrations for different shaped pores using the pore size that returns the largest enhancement in thermopower for that carrier concentration. The cubic pores show slightly better performance and enhanced the Seebeck coefficient up to 15\% at low carrier concentrations and $\sim 60\%$ at high carrier concentrations while the enhancement is limited to 12\% and 30\% at low and high carrier concentration regimes, respectively for spherical pores. Note than the largest enhancement in Seebeck generally takes place at medium-level concentrations ($\sim 10^{20}\ \mathrm{1/cm^3}$), e.g., the largest fractional enhancement in Seebeck for cylindrical pore, occurs at $\sim 1.5 \times 10^{20}\ \mathrm{1/cm^3}$ concentration. At higher temperatures, the scope for fractional improvement in Seebeck is not as dramatic but the magnitude of enhancement is still larger than $20\ \mu V/K$ — see figure \ref{fig:fig8} in appendix B. The right-hand plot in figure \ref{fig:fig3} shows $\Delta_1$ — the average energy of the current carrying electrons — as a function of pore size at $1.2 \times 10^{20}\ \mathrm{1/cm^3}$. The green line shows $\Delta_1$ in bulk Si and is equal to 126 meV. The cubic pores provide the largest $\Delta_1$ enhancement followed by spherical pores ($\Delta_1 \approx 138\ \mathrm{meV}$). This corresponds to $\sim 9.5\%$ enhancement in thermopower due to energy electron filtering — with the optimal characteristic length of the pores being 0.75 nm and 1.00 nm, respectively. The key message from the plot of $\Delta_1$ in figure \ref{fig:fig3} is that, at high carrier concentrations, there is only little additional return on the effort required to make pores have particular geometry — most of the benefit comes from making the pores small. This means that as a design strategy for thermopower enhancement one should seek to create pores of any shape, but to make them as small as possible. We note that the largest enhancement in $S$ does not necessarily provide the maximum power factor. For the best PF performance, the countervailing response of enhancement in $S$ and reduction in $\sigma$ should be considered simultaneously.

\begin{figure}[h]
\centering
\includegraphics[width=0.5\textwidth]{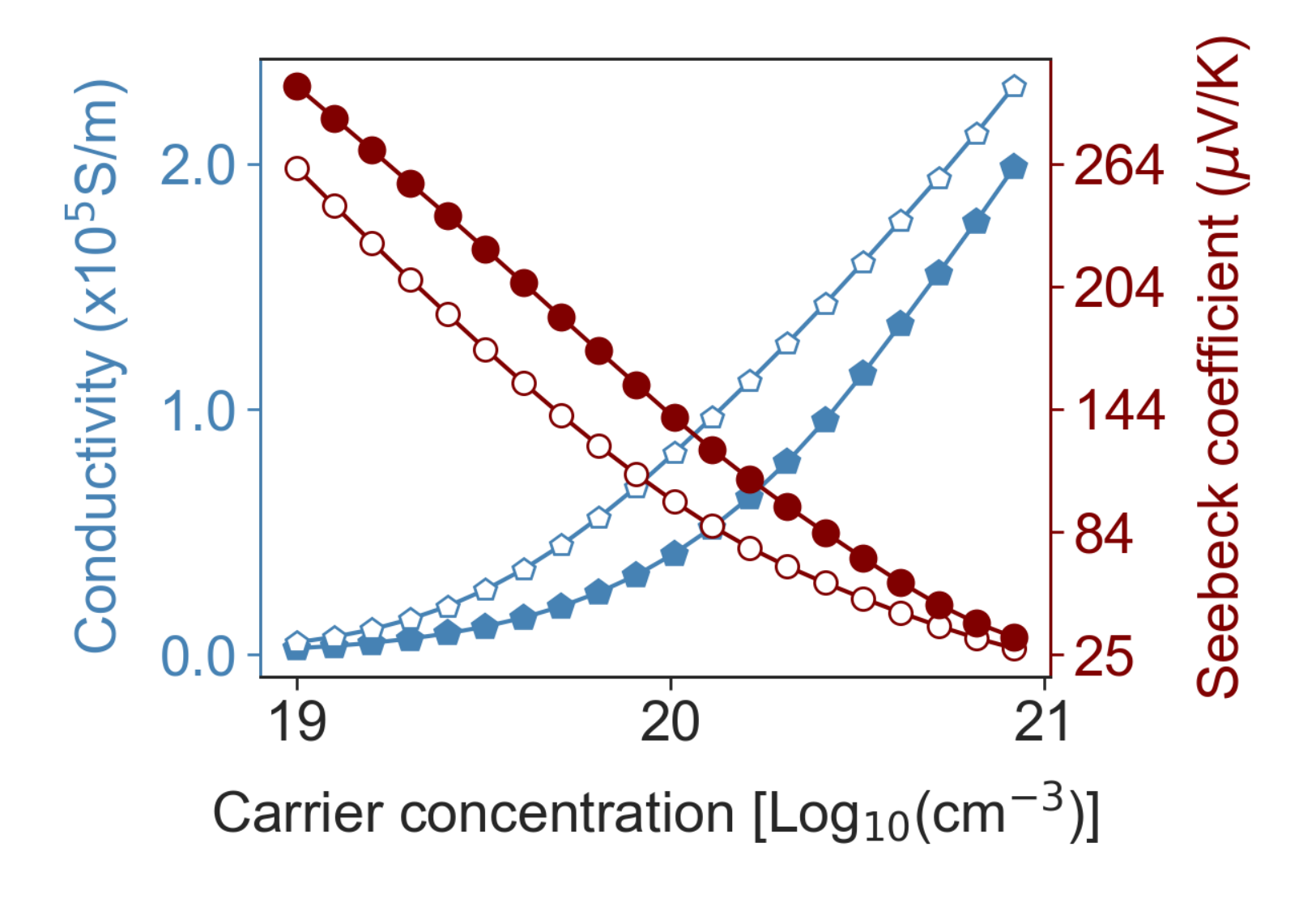}
\caption{Electrical conductivity (blue) and thermopower (red) vs carrier concentration at 300 K. Open symbols are for bulk \textit{n}-type Si. Solid symbols are for Si containing 0.05 spherical porosity with the characteristic length of 1.67 nm. The maximum enhancement in PF happens at $3.2 \times 10^{20}\ \mathrm{1/cm^3}$.}
\label{fig:fig4}
\end{figure}

Figure \ref{fig:fig4} shows the electrical conductivity (pentagons) and Seebeck coefficient (circles) in bulk with open marker and in 5\% of spherical porosity with characteristic length of 1.67 nm with close marker. The Seebeck coefficient shows around 40\% increase at $1.8 \times 10^{20}\ \mathrm{1/cm^3}$ while the largest degradation in conductivity is about 55\% and happened at low carrier population of $3.2 \times 10^{19}\ \mathrm{1/cm^3}$. The enhancement in Seebeck ($S^2$ in PF) offsets the reduction in conductivity for carrier populations beyond $10^{20}\ \mathrm{1/cm^3}$. The maximum enhancement of PF is about 35\% and takes place at $3.2 \times 10^{20}\ \mathrm{1/cm^3}$ concentration.

\begin{figure}[h]
\centering
\includegraphics[width=1\textwidth]{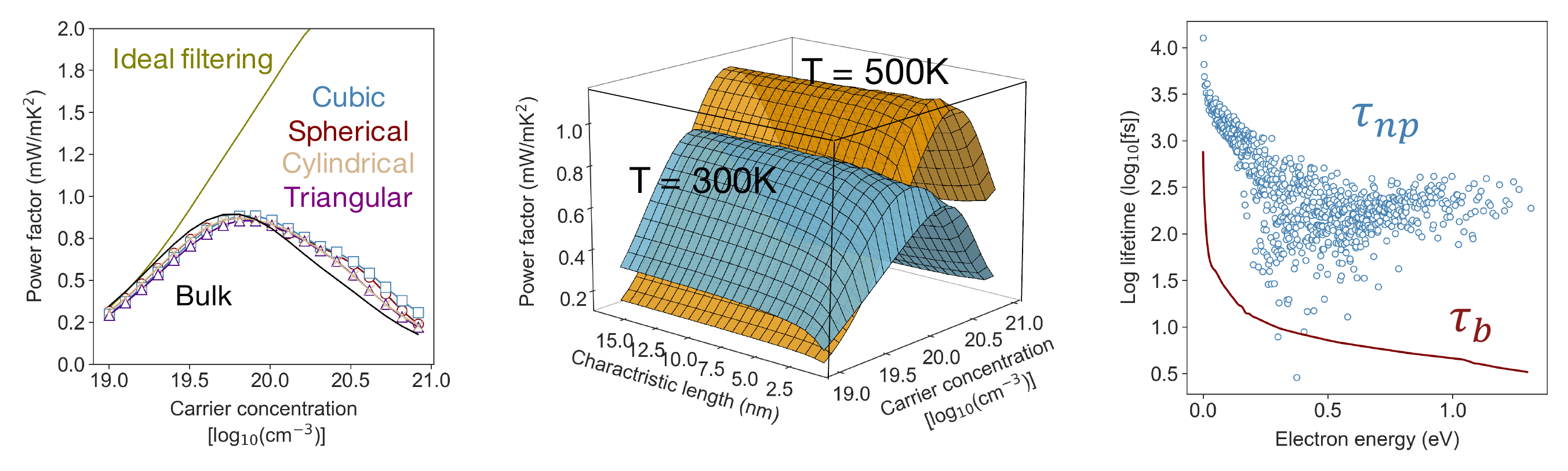}
\caption{(Left) Variation in PF with carrier concentrations for pores with different shapes. The PF in bulk Si is marked with open circle in black. The largest enhancement with the ideal filtering is plotted in green. The pores with different shapes demonstrate similar behavior with reasonably good power performance. At room temperature, the maximum PF in porous structure always happens in higher carrier concentration than the carrier concentration that the bulk Si shows the best performance. (Middle) The variation in PF with pore length and carrier concentration in spherical pores at 300 and 500 K. (Right) Electron-pore lifetime for extended cylindrical pores at 300 K is plotted in blue. The characteristic length is 3.3 nm. The electron lifetime due to combination of phonons and ions is plotted in red. This is the dominant scattering term. For extended pores with low porosity, the Seebeck is similar to the bulk and the electrical conductivity is $(1-\varphi)$ of the bulk counterpart.}
\label{fig:fig5}
\end{figure}

The left-hand panel in figure \ref{fig:fig5} shows the model prediction for the variation in largest achievable TE power factor with carrier concentration in Si based porous materials with optimal characteristic lengths at 300 K. The best power factor performance using the ideal filtering model is plotted in green. The power factor in bulk Si is plotted in black. In the narrow carrier concentration window with the highest power factor, bulk Si shows slightly better performance. The large energy difference between the conduction band edge in the Si and the vacuum level in the pore (about 4.15 eV electron affinity of bulk Si \cite{burton1976temperature}) causes strong electron scattering with countervailing response of reduction in the electrical conductivity and enhancement of Seebeck coefficient that cancel out each other, leading to an overall unchanged PF value. The maximum PF in porous structures takes place at carrier concentrations higher than the optimal carrier concentration in bulk Si. This is a key insight for the design of thermoelectrics at room temperature: If one is planning to engineer porous thermoelectrics to reduce phonon conduction, then one should also plan to increase the carrier concentration above the optimal level for the bulk semiconductor. In the Si model the maximum power factor of porous materials takes place at $8 \times 10^{19}\ \mathrm{1/cm^3}$ carrier concentration and is slightly less than the maximum power factor in bulk Si occurs at $6.3 \times 10^{19}\ \mathrm{1/cm^3}$, i.e., $\sim 25\%$ increase in doing concentration is needed for the best performance in porous Si. The power factor in Si with spherical pores at 300 K and 500 K for different characteristic length and concentration is shown in the central panel in figure \ref{fig:fig5}. At higher temperature of 500 K a less extreme increase in carrier concentration is needed to recuperate the power factor, and the recovery is larger. As an example, the maximum power factor of spherical pores at this temperature takes place at $1.6 \times 10^{20}\ \mathrm{1/cm^3}$ carrier concentration and is slightly larger than the maximum power factor in bulk Si that takes place at $1.26 \times 10^{20}\ \mathrm{1/cm^3}$ carrier concentration ($\sim 20\%$ increase in doing concentration).

To complete the model of electron-pore interaction, we considered extended cylindrical with infinite length (system size) oriented along the (001) crystal axis (the z-direction is our reference system) in a P-doped silicon slab. The thermal and electrical properties of such porous Si films have been studied in \cite{romano2014toward,de2019large,de2020heat}, usually with the assumption that electron scattering is the same as that in bulk Si \cite{lee2008nanoporous}. The electron lifetime of P-doped Si with extended cylindrical pores with 10 nm radius (3.3 nm characteristic length) and 0.05 porosity at 300 K is depicted in the right-hand panel of figure \ref{fig:fig5}. This plot shows that electron-phonon and electron-ion scattering is dominant over the pore scattering by one to two orders of magnitude. The dramatic reduction in the rate of electron scattering from discrete to extended pores is due to the limited number of states that are available to accept scattered electrons. The analytic expressions for the scattering matrix elements for these extended pores are given in the appendix A. In extended pores, scattering is only possible into states with the same component of wave vector along the pore axis. This condition, combined with the isoenergetic constraint, reduces the scattering integral to an elliptical line, drastically reducing the number of states that can participate in scattering, and means that the extended pores cause no change in the electron momentum along the axis of the pores. This result strengthens the assumption made in prior works \cite{lee2008nanoporous,lee2009thermoelectric} that extended pores do not change the electron lifetime and thus the Seebeck coefficient of 2D nanoporous Si is the same as the bulk Si and electrical conductivity and power factor in 2D porous Si with translational invariance vertical to the simulation plane are $(1-\varphi)$ of their bulk Si counterparts.

\begin{figure}[h]
\centering
\includegraphics[width=0.5\textwidth]{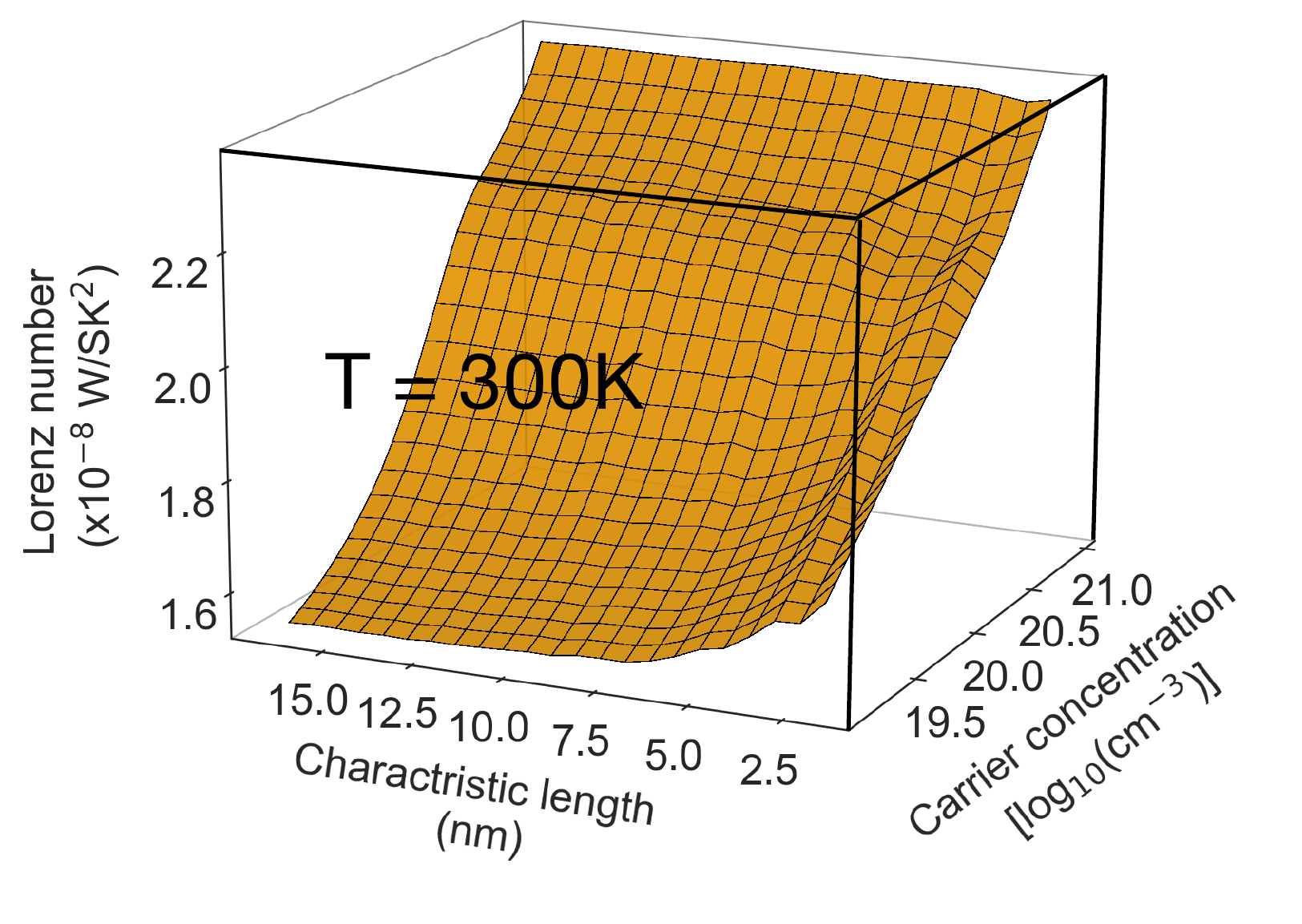}
\caption{Variation of Lorenz number by the carrier concentration and characteristic length of cylindrical pores at 300 K.}
\label{fig:fig6}
\end{figure}

When advocating for increased carrier concentration in thermoelectrics, it is important to determine if this will cause a significant increase to the denominator of ZT. Hence, we finish our examination of the effect of pores on the room temperature electrical transport coefficients by briefly discussing the electronic thermal conductivity ($\kappa_e$). The $\kappa_e$ is related to $\sigma$ by Wiedemann Franz law as $\kappa_e = L T \sigma$. Here $L$ is the Lorenz number that conventionally varies from $2 \times (k_B/e)^2 \approx 1.48 \times 10^{-8}\ \mathrm{(V^2/K^2)}$ up to $\pi^2/3 \times (k_B/e)^2 \approx 2.44 \times 10^{-8}\ \mathrm{(V^2/K^2)}$ for low carrier concentration and degenerate (free electron) limit, respectively \cite{kim2015characterization}. Lorenz number is related to the moments of the charge carriers, $\Delta_n$, through $L = 1/(eT)^2  (\Delta_2-\Delta_1^2)$. In bulk Si, the Lorenz number varies monotonically from $1.53 \times 10^{-8}\ \mathrm{(V^2/K^2)}$ at $10^{19}\ \mathrm{1/cm^3}$ to $2.39 \times 10^{-8}\ \mathrm{(V^2/K^2)}$ at $10^{21}\ \mathrm{1/cm^3}$. Figure \ref{fig:fig6} shows the variation of Lorenz number with characteristic length and carrier population in porous Si with cylindrical pores at 300 K — see figure \ref{fig:fig8} in appendix B for Lorenz number in nanoporous Si with different pore shapes. While Lorenz number varies considerably with the carrier concentration it has limited dependency on pores’ characteristic length especially in high carrier concentration regime. Figure \ref{fig:fig9} in appendix B shows the largest and the lowest changes in Lorenz number for the pores with different shapes and sizes in this study. The largest enhancements are shown with solid markers and the lowest values of the Lorenz number are shown with open markers. The Lorenz number in bulk silicon is plotted in solid black. Similar to the Seebeck coefficient, cubic pores show the largest impact on Lorenz number followed by spherical pores. Although this result is interesting, the overall impact of these changes in Lorenz number for the optimization of thermoelectric ZT will be minimal. The objective of adding porosity is to lower the lattice thermal conductivity, and prior works by Romano \cite{romano2014toward} and others \cite{hsieh2012thermal} have shown that the lattice thermal conductivity in nanoporous Si with the geometries modeled here can be as low as $\sim 30\ \mathrm{W/m/K}$ at room temperature.  In bulk room temperature Si with the carrier concentration tuned to optimize ZT, the electronic thermal conductivity is $\sim 0.3\ \mathrm{W/m/K}$ — still two orders of magnitude lower than the lattice conductivity.

\begin{figure}[h]
\centering
\includegraphics[width=0.5\textwidth]{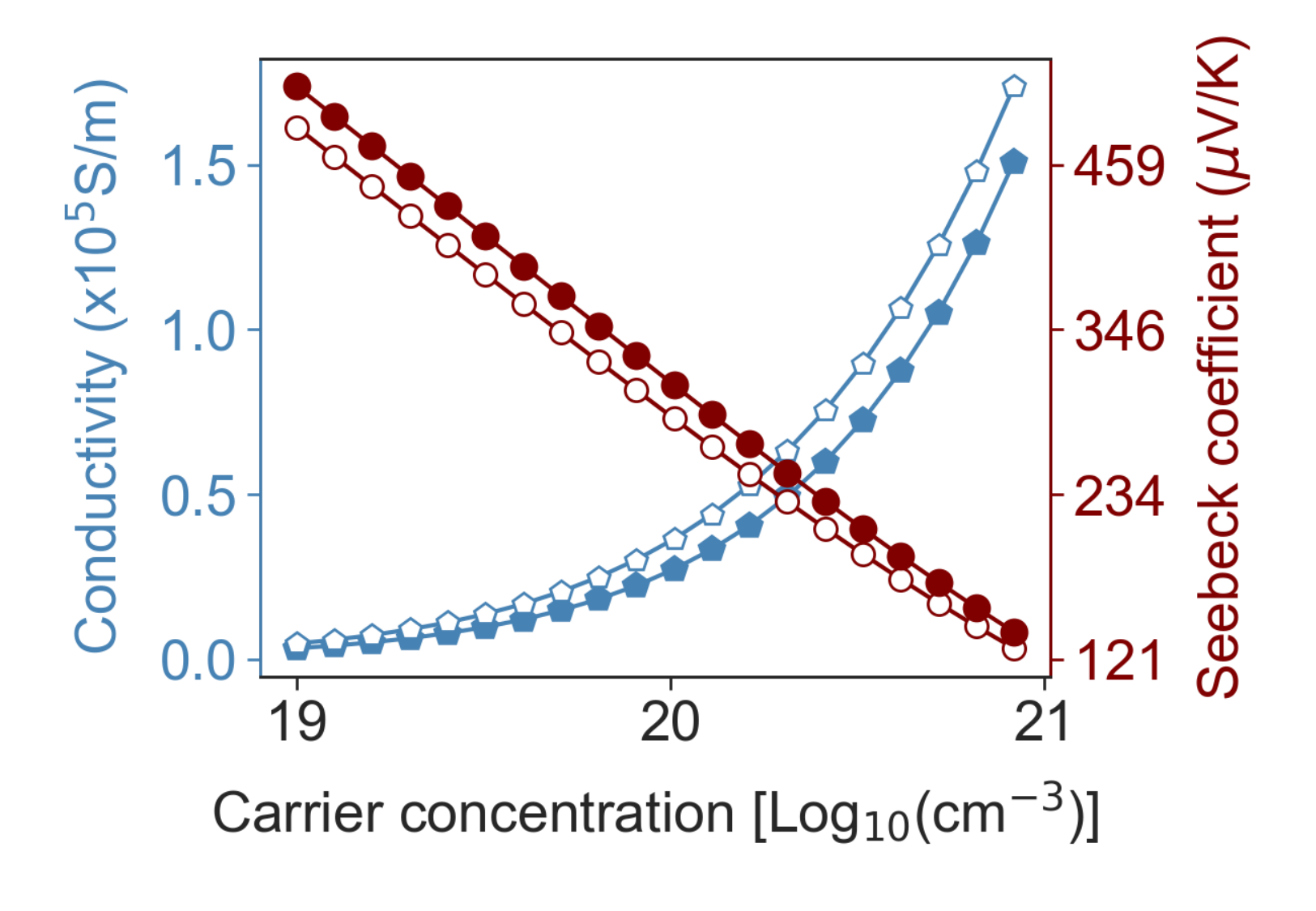}
\caption{Electrical conductivity (blue) and thermopower (red) vs carrier concentration at 1300 K. Open symbols are for bulk \textit{n}-type Si. Solid symbols are for Si containing 0.05 spherical porosity with the characteristic length of 1.67 nm. The scope of electron filtering is limited at high temperatures.}
\label{fig:fig7}
\end{figure}

To complete the discussion on the effect of pores on Si based porous materials we computed the transport coefficients at 1300 K (figure \ref{fig:fig7}). The bulk properties are shown with open markers and the properties for 5\% spherical pores are plotted with solid markers. Although the magnitude of power factor is larger at high temperatures the scope of PF enhancement via electron filtering is limited. Therefore, the power factor in bulk is larger than the porous Si in all ranges of carrier concentrations. The maximum PF in both bulk and porous Si takes place at $3.2 \times 10^{20}\ \mathrm{1/cm^3}$ and the enhancement in Seebeck because of filtering effect mitigates up to 95\% of the PF in bulk.

\section{CONCLUSION}

To summarize, we have used a semiclassical model to elucidate the detrimental effect that porosity has on the electrical transport properties of thermoelectric, and to devise design strategies to mitigate them. We have shown that while extended pores have little effect on electron scattering, scattering from compact pores provides an electron filtering effect that increases the Seebeck coefficient. This effect becomes more pronounced for smaller pores but is relatively insensitive to the pore geometry. We find that to take full advantage of this effect to mitigate the degradation that pores cause to thermoelectric PF one should increase the carrier concentration above the optimal level for monolithic semiconductor. In this case one can recuperate as much as 95\% of the lost thermoelectric PF because of the pores. While we have focused in particular a semiclassical model of P-doped Si, as this is a model that has been experimentally validated, the findings should be transferable to other semiconductors systems both \textit{n}- and \textit{p}- type and with either direct or indirect band gaps. While there is much focus currently on designing porosity in thermoelectrics to dramatically impede phonon transport, the results presented here form a complementary design principle for optimizing the electrical transport properties in such devices. The electrical transport properties in the numerator of ZT are less sensitive to pore shape than the phonon transport properties in the denominator.  This means that a good strategy for designing nanoporous thermoelectrics that maximize ZT is to first focus on optimizing pore size and morphology to maximize phonon scattering, and then to adjust the carrier concentration to mitigate the damage to the electrical transport properties. 

\section{DATA AVAILABILITY}
The data that support the findings of this study are available from the corresponding author upon reasonable request.
\section{CONFLICT OF INTEREST}
The authors declare no competing financial interest.
\section{APPENDIX}

\subsection{Appendix A: Electron Matrix Elements of Pores with Different Shapes}
The electron matrix element shows the strength of the coupling between initial and final wavefunctions and the number of ways the transmission may happen. For the Bloch waves, the matrix element relies on the shape of the scattering potential. Here we present the full expression of matrix elements for pores with cubic, spherical, triangular and cylindrical shapes followed by matrix elements for different shaped extended pores.

For cubic pores with finite lengths of $l_x$, $l_y$ and $l_z$ along $x$, $y$ and $z$ direction, respectively, electron matrix element describes as
\begin{equation}
    M_{kk'}= 8U_o \left( \frac{\sin\left(\frac{l_xq_x}{2}\right)\sin\left(\frac{l_yq_y}{2}\right)\sin\left(\frac{l_zq_z}{2}\right)}{q_xq_yq_z}\right)
\end{equation}
In this equation, $q=k-k'$, and $q_x=q.\hat{i}$, $q_y=q.\hat{j}$, $q_z=q.\hat{k}$ are the projection of $q$ on Cartesian axes. In prism with isosceles triangle base, matrix element defines as
\begin{equation}
    M_{kk'}= -4U_ol_y \left( \frac{l_xq_x-2l_yq_y -2l_xq_xe^{i\left(\frac{l_xq_x}{2}+l_yq_y\right)}+l_xq_xe^{i(l_xq_x)} + 2l_yq_ye^{il_xq_x}}{l_x^2q_x^3-4l_y^2q_xq_y^2}\right) \left( \frac{\sin\left( \frac{l_zq_z}{2}\right)}{q_z}\right)
\end{equation}
In this equation, $l_x$ and $l_y$ are the length and height of the triangle, respectively and $l_z$ is the height of the prism. For cylindrical potential, we have
\begin{equation}
    M_{kk'}= 4\pi r_o U_o \left(\frac{J_1\left(r_oq_r\right)}{q_r}\right)\left(\frac{\sin\left(\frac{l_zq_z}{2}\right)}{q_z}\right)
\end{equation}
In this equation, $q_r=\sqrt(q_x^2+q_y^2)$, $r_o$ is the radius of the base circle, $l_z$ is the height of the cylinder and $J_1$ is the first order Bessel function of the first kind. Electron coupling matrix element for spherical potential is defined as
\begin{equation}
    M_{kk'}= \frac{4\pi U_o}{q^2} \left(\frac{1}{q}\sin(r_oq)-r_o\cos(r_oq)\right)
\end{equation}
where, $q$ is the magnitude of $\mathbf{q}$ and $r_o$ is the radius of pores. For cubic pores with infinite length (system size) along $z$, the matrix element is defined as
\begin{equation}
    M_{kk'}= 4 U_o l_z \left( \frac{\sin\left(\frac{l_xq_x}{2}\right)\sin\left(\frac{l_yq_y}{2}\right)}{q_xq_y}\right)\delta(q_z)
\end{equation}
For the cylindrical pore with infinite height, we have
\begin{equation}
    M_{kk'}= 2\pi r_o U_o l_z \left(\frac{J_1\left(r_oq_r\right)}{q_r}\right)\delta(q_z)
\end{equation}
For the isosceles triangular prism with infinite height matrix element is described as 
\begin{equation}
    M_{kk'}= -2U_ol_yl_z \left( \frac{l_xq_x-2l_yq_y -2l_xq_xe^{i\left(\frac{l_xq_x}{2}+l_yq_y\right)}+l_xq_xe^{i(l_xq_x)} + 2l_yq_ye^{il_xq_x}}{l_x^2q_x^3-4l_y^2q_xq_y^2}\right) \delta(q_z)
\end{equation}

\subsection{Appendix B: The Effect of Nanopores on Lorenz Number at Room Temperature}

Figure \ref{fig:fig8} shows changes in Lorenz number with carrier concentrations for pores with different shapes and sizes at room temperature for 5\% porosity. Similar to the cylindrical pore, the Lorenz number is larger for smaller pores and the pores size has stronger effect at regimes with lower carrier populations. The maximum and minimum computed Lorenz number and the bulk value of the Lorenz number are plotted in figure \ref{fig:fig9}.

\subsection{Appendix C: The Effect of Pores on Electrical Properties of Silicon-Based Nanoporous at High Temperatures}

Figure \ref{fig:fig10} shows the variation of highest Seebeck (thermopower) and power factor modeled in this study with carrier concentration for pores with different shapes at 500 K and 1300 K. The bulk properties are shown in solid black lines.

\bibliographystyle{unsrt}  
\bibliography{references} 

\newpage

\section*{Figures}


\begin{figure}[h]
\centering
\includegraphics[width=1\textwidth]{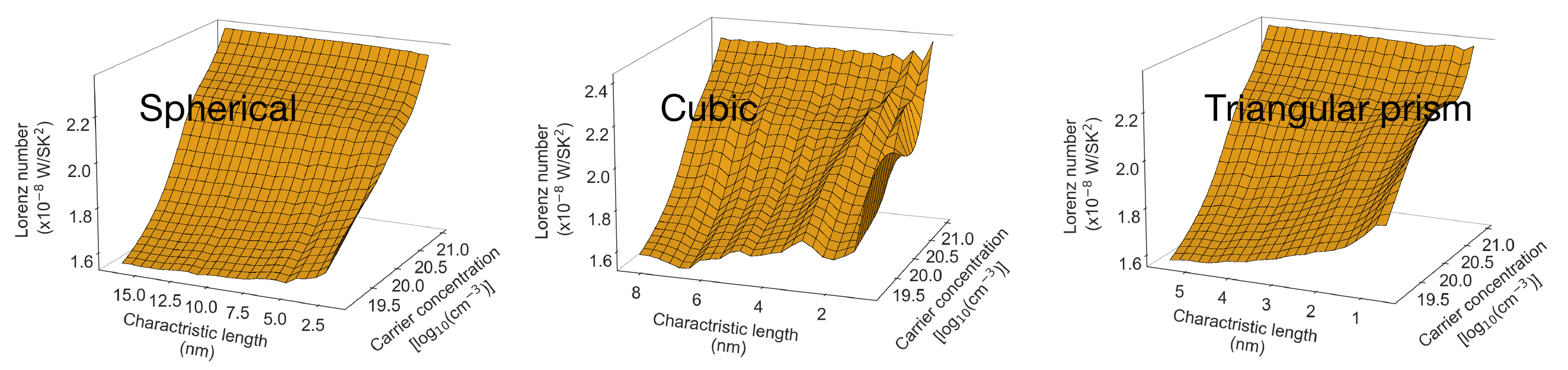}
\caption{Variation of Lorenz number with carrier concentrations for spherical, cubic, and triangular prism pores with different characteristic lengths at 300 K.}
\label{fig:fig8}
\end{figure}


\begin{figure}[h]
\centering
\includegraphics[width=0.5\textwidth]{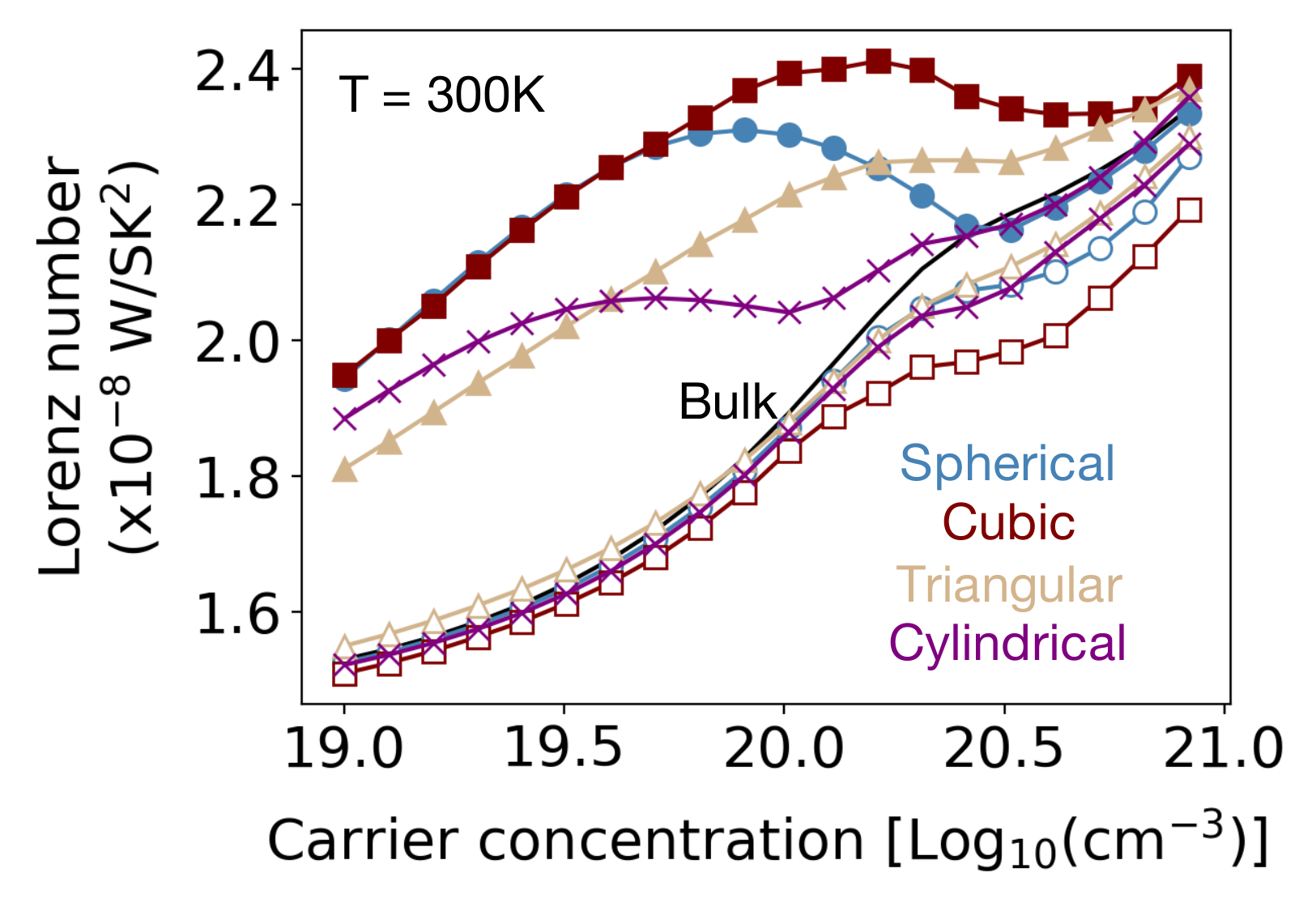}
\caption{Variation of Lorenz number with carrier concentrations in bulk Si is plotted in solid black. The highest and lowest values of the Lorenz number for the pores modeled in this study are shown with solid and open markers, respectively.}
\label{fig:fig9}
\end{figure}


\begin{figure}[h]
\centering
\includegraphics[width=1\textwidth]{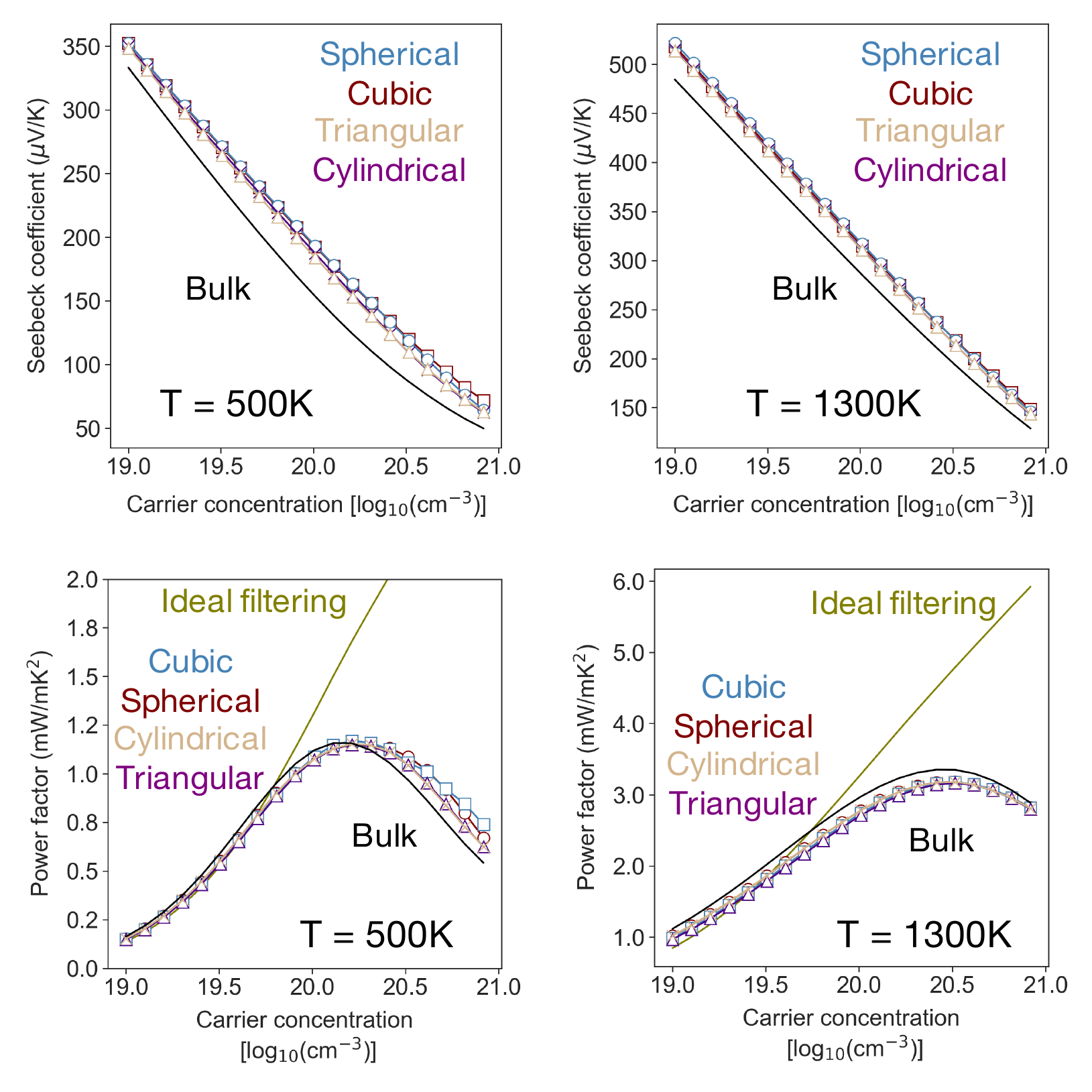}
\caption{(Top) Maximum enhancement is thermopower for pores with different shapes at 500 K (left) and 1300 K (right). (Bottom) Comparison of the largest achievable PF with the filtering effect using pores with different shapes. The largest achievable PF with the filtering effect using an ideal model is plotted in green. The PF in bulk Si is shown in black. The left panel is at 500 K, and the right panel is at 1300 K.}
\label{fig:fig10}
\end{figure}

\end{document}